\documentclass[a4paper,fleqn]{cas-dc}

\usepackage[authoryear,longnamesfirst]{natbib}
\usepackage{amsmath,amssymb}
\usepackage{array,booktabs,tabularx}
\usepackage{graphicx}
\usepackage{subcaption}
\begin{document}

\shorttitle{Differentiable Gradient Enhanced Damage}
\shortauthors{Wilkinson et al.}

\title[mode=title]{A Differentiable Framework for Gradient Enhanced Damage with Physics-Augmented Neural Networks in JAX-FEM}

\author[1]{Mark Wilkinson}
\author[2]{Amirhossein Amiri-Hezaveh}
\author[2]{Adrian Buganza Tepole}

\affiliation[1]{organization={Purdue University}}
\affiliation[2]{organization={Columbia University}}
\begin{abstract}
Soft materials such as rubbers, hydrogels, and biological tissues undergo damage in the form of stiffness degradation without apparent changes in their stress-free geometry. Accurate simulation of this behavior is critical in applications ranging from soft robotics to the design of medical devices, yet two persistent challenges are the difficulty of constructing flexible, thermodynamically consistent constitutive models, and the mesh dependence of finite element solutions caused by strain softening. Here we address both challenges simultaneously by combining physics-augmented neural network constitutive models with a gradient-enhanced damage formulation implemented within the differentiable finite element framework JAX-FEM. The elastic strain energy and the damage yield function are each parameterized by input-convex neural networks (ICNNs), which enforce polyconvexity and satisfaction of the Clausius--Duhem inequality by design. The gradient-enhanced formulation introduces a non-local damage field governed by an additional partial differential equation, regularizing the spatial distribution of damage and eliminating mesh dependence. The implementation is validated through local stress-strain fits, single-element parametric studies, a mesh and solution strategy study for a uniform deformation case, and a notched plate simulation. The results demonstrate that the proposed framework enables flexible, data-driven, mesh-independent damage simulation for a broad class of soft materials. We anticipate that the open-source implementation will lower the barrier to adopting physics-augmented neural network constitutive models.
\end{abstract}
\maketitle
\section{Introduction}

Soft materials undergo damage in the form of stiffness degradation without apparent changes in their stress-free geometry \citep{li2016damage,holzapfel2020damage}. This phenomenon was first observed in rubber in what is known as the Mullins effect, and has since been identified in materials such as biological tissues and hydrogels \citep{mullins1948effect,webber2007large}. Its distinguishing feature is a hysteretic stress-strain response: upon first loading, the material follows an initial curve, but on unloading the stress lies below the loading path. Subsequent reloading then traces the unloading path until the previously attained maximum deformation, beyond which dissipation resumes and permanent stiffness degradation accumulates. At sufficiently large deformations, the stiffness reduction is severe enough to produce softening, i.e. a negative tangent stiffness, which can rapidly drive failure under a sustained load.

Accurate modeling of damage in soft materials is essential across a range of classical and emerging applications. Tough hydrogels, for instance, are increasingly studied for their energy dissipation characteristics \citep{zhao2017designing}. Soft materials produced by additive manufacturing are subjected to extreme deformations in applications such as soft robotics, where reliable damage predictions are critical \citep{gomez20213d}. Biological tissues similarly exhibit damage at large deformations, and faithful models of this behavior are needed to guide the design of medical devices and interventions \citep{rausch2017modeling}.

A standard modeling framework is to consider an internal variable for damage together with a degradation function that reduces the stored elastic energy. A widely adopted formulation, due to \cite{simo1987strain}, defines the degraded free energy as $(1-d)\psi$, where $d \in [0,1]$ is the damage variable and $\psi$ is the undamaged strain energy density; $d = 0$ corresponds to the intact material and $d = 1$ to complete loss of load-carrying capacity. Damage evolution is governed by a yield surface that tracks the maximum deformation in the history of loading. Damage accumulates only when this surface is reached. The second law of thermodynamics requires that the dissipation should remain non-negative throughout. This is enforced through associative flow rules derived from the principle of maximum dissipation, with a yield function $\mathcal{G}$ that has appropriate monotonicity properties with respect to the thermodynamic conjugate of $d$ \citep{simo1987strain,tac2024data}. Alternative approaches include the pseudo-elastic framework of \cite{ogden1999pseudo}, the metric-based formulation of \cite{menzel2001theoretical}, micromechanics-based approaches \citep{zhan2023general}, among others \citep{ricker2021comparison}. In summary, the continuum damage model of a material can be specified by a strain energy function and a dissipation potential.

Closed-form choices for these two functions are material-specific. In many cases, analytical expressions do not adequately capture experimental data across a broad range of deformations. This has motivated a growing body of work in scientific machine learning aimed at data-driven constitutive modeling. Approaches such as Constitutive Artificial Neural Networks (CANN) \citep{linden2023neural,holthusen2024theory} and EUCLID~\citep{flaschel2023automated,thakolkaran2022nn} pursue interpretable model identification, while physics-augmented neural networks offer more flexibility at the cost of interpretability but still enforce physical constraints directly by construction~\citep{kalina2026physics,tepole2025polyconvex}. For the strain energy, key requirements include objectivity, material frame indifference, and polyconvexity. Input-convex neural networks (ICNNs) have been used to satisfy these constraints by design~\citep{amos2017input}. For the dissipation potential, we have previously developed monotonic neural network architectures, including formulations based on neural ordinary differential equations (NODEs), that guarantee satisfaction of the dissipation inequality by design~\citep{tac2022data}. Despite this flexibility, neural network constitutive models have seen limited adoption in finite element simulations to date.

A further challenge specific to damage modeling in finite elements is the well-known mesh-dependence induced by strain softening. Once an element reaches the yield surface and begins to soften, it localizes all further dissipation, shielding adjacent elements from additional damage degradation. The resulting failure pattern is governed by the mesh rather than any intrinsic material length scale, which is physically unrealistic. Non-local damage formulations resolve this by coupling the local constitutive response, which depends only on the deformation history at a single material point, with a global smoothing field that regularizes the spatial propagation of damage and introduces a characteristic length scale \citep{he2019gradient}. Several such formulations exist with some approaches involving either integral smoothening of the local field \citep{ferreira2017modeling,suarez2024nonlocal}, or the introduction of a gradient-based non-local regularization. The former involves a volumetric smoothening applied directly on the local field, while the latter introduces a separate gradient coupling within the free-energy \citep{critical_gonalves_2025,waffenschmidt_gradient-enhanced_2014,efficient_seupel_2018}. 

In this work, we combine physics-augmented neural network constitutive models with a gradient-enhanced damage formulation, following the approach of \cite{ostwald_implementation_2019}. Physics-augmented neural networks have been implemented in several languages and libraries, with JAX and PyTorch being particularly prominent in the scientific machine learning community. A key feature of JAX is its support for differentiable programming: by tracing the computational graph of a given operation, JAX enables automatic differentiation via reverse-mode chain rule without requiring analytical gradient derivations. This capability has been exploited in the development of JAX-FEM~\citep{xue2023jax}, a differentiable finite element framework. The direct compatibility between JAX-FEM and our JAX-based neural network models makes it a natural platform for the present work. We therefore present the formulation and implementation of a gradient-enhanced damage model with physics-augmented neural network constitutive laws within JAX-FEM, enabling, for the first time, flexible data-driven damage simulation with mesh-objective finite element solutions.

\section{Local and Non-local Damage Formulations}
For the development of the damage formulation, we first introduce the basic kinematic quantities and their role in the free energy function. The free energy is composed of a strain energy component plus additional contributions from the non-local damage field. This free energy is then leveraged within a variational form, reducing to two separate PDEs for the displacement and non-local damage fields respectively. In this section we keep the constitutive models generic, with subsequent specialization to closed-form or data-driven frameworks.  

\subsection{Variational Statement}\label{sec:damage_anlytic}
We first propose the existence of some free energy function $\Psi_{\mathrm{int}}$ for a given system, initially defined in a reference configuration $\mathcal{B}_0$, with motion $\varphi(\mathbf{X},t):\mathcal{B}_0\rightarrow\mathcal{B}_t$ to a deformed configuration $\mathcal{B}_t$. Additionally, we introduce the existence of both a local internal variable $\kappa$ and a non-local scalar field $\phi$ associated with the body. Taking the deformation gradient $\mathbf{F}=\nabla_X\varphi$, the free energy is assumed to be an additive split of local and non-local contributions,

\begin{equation}
\Psi_{\mathrm{int}}(\mathbf{F},\phi,\nabla_X\phi,\kappa)
= \psi_{\mathrm{loc}}(\mathbf{F},\kappa)
+ \psi_{\mathrm{nloc}}(\mathbf{F},\phi,\nabla_X\phi,\kappa).
\label{eq:internal_energy_split}
\end{equation}
This internal free energy function can be further specified by associating the local contribution to a strain energy density function, $\psi_e(\mathbf{C})$, scaled by a local damage function $f_d(\kappa)$, and by splitting the non-local contributions into terms involving the field $\phi$ only and those involving its gradient $\nabla_X \phi$,

\begin{equation}
\begin{aligned}
\Psi_{\mathrm{int}}(\mathbf{F},\phi,\nabla_X \phi,\kappa) 
  &= f_d(\kappa)\,\psi_e(\mathbf{C}) \\
  &\quad + \psi_{\mathrm{nloc}}^{\mathrm{grad}}(\nabla_X\phi;\mathbf{F}) \\
  &\quad + \psi_{\mathrm{nloc}}^{\mathrm{plty}}(\phi,\kappa) ,
\end{aligned}
\label{eq:internal_energy}
\end{equation}
with $\mathbf{C}=\mathbf{F}^{\mathsf{T}}\mathbf{F}$ being the right Cauchy-Green tensor. The damage function is assumed to be a monotonically decreasing function of $\kappa$ satisfying $f_d(0)=1$, $0<f_d(\kappa)\le 1$, and $f_d'(\kappa)\le 0$, with $\kappa=0$ corresponding to the undamaged state. The remaining terms, $\psi_{\mathrm{nloc}}^{\mathrm{grad}}(\nabla_X\phi;\mathbf{F})$ and $\psi_{\mathrm{nloc}}^{\mathrm{plty}}(\phi,\kappa)$, are the gradient and penalty contributions to the non-local damage field to enforce smoothness with a given length-scale and to couple the non-local field $\phi$ to the local damage evolution described by $\kappa$.

The free energy is then used to propose a variational statement for the total potential energy of the system. The total potential energy is given by the difference of the internal energy and the work done by the body forces $\bar{\mathbf{B}}$ and surface tractions $\bar{\mathbf{T}}$ over the domain.

\begin{equation}
\begin{aligned}
\Pi(\varphi,\phi,\kappa)
={} & \int_{\mathcal{B}_0}\Psi_{\mathrm{int}}(\mathbf{F},\phi,\nabla_X \phi,\kappa)\,\mathrm{d}V \\
   & - \int_{\mathcal{B}_0}\bar{\mathbf{B}}\cdot\varphi\,\mathrm{d}V - \int_{\partial\mathcal{B}_0}\bar{\mathbf{T}}\cdot\varphi\,\mathrm{d}A.
\end{aligned}
\label{eq:total_potential_energy}
\end{equation}
The boundary value problem is governed by the principal of minimum potential energy. To obtain the coupled boundary value problem, the stationarity condition $\delta \Pi=0$ is applied to the total potential energy to find the minimum. From inspection, the first variation decomposed into the separate inputs of the potential as

\begin{equation}
\begin{aligned}
\delta \Pi(\boldsymbol{\varphi}, \phi, \kappa) 
&= 
\frac{\partial \Pi}{\partial (\nabla_X \boldsymbol{\varphi})} : \nabla_X \delta \boldsymbol{\varphi}
+ \frac{\partial \Pi}{\partial \boldsymbol{\varphi}} \cdot \delta \boldsymbol{\varphi} \\
&\quad
+ \frac{\partial \Pi}{\partial (\nabla_X \phi)} \cdot \nabla_X \delta \phi
+ \frac{\partial \Pi}{\partial \phi} \, \delta \phi
\\
&\quad = 0.
\end{aligned}
\end{equation}
where we have used $\delta \nabla_X( \boldsymbol{\varphi}) = \nabla_X \delta \boldsymbol{\varphi}$ and $\delta \nabla_X( \phi) = \nabla_X \delta \phi$ to simplify the variation of the gradients of the fields of interest, $\varphi$ and $\phi$.  Note that the local damage $\kappa$ is not part of the variational statement. Instead, the local damage is an internal variable that satisfies an auxiliary evolution equation to drive energy dissipation. The evolution equation for $\kappa$ is constrained by the second law of thermodynamics expressed through the Clausius-Duhem inequality. Examples of the evolution equation for $\kappa$ are addressed below. We also remark that the variations of the global field, $\delta \varphi$ and $\delta \phi$ are not completely arbitrary. Rather, $\delta \varphi$ and $\delta \phi$ vanish on the boundary of the domain with Dirichlet boundary conditions for $\varphi$ and $\phi$ respectively.

The variation of total potential decomposes into $\delta \Pi = \delta \Pi_{\varphi} + \delta \Pi_{\phi} = 0$, where both terms must vanish. The stationary conditions become

\begin{equation}
\begin{aligned}
\delta_{\varphi}\Pi
={}&\int_{\mathcal{B}_0}\frac{\partial \Psi_{\mathrm{int}}}{\partial \mathbf{F}}:\nabla_X\delta\varphi\,\mathrm{d}V \\
&-\int_{\mathcal{B}_0}\bar{\mathbf{B}}\cdot\delta\varphi\,\mathrm{d}V
-\int_{\partial\mathcal{B}_0}\bar{\mathbf{T}}\cdot\delta\varphi\,\mathrm{d}A =0,
\end{aligned}
\end{equation}
while the weak form associated with the non-local damage field is

\begin{equation}
\begin{aligned}
\delta_{\phi}\Pi
={}&\int_{\mathcal{B}_0}
\frac{\partial \Psi_{\mathrm{int}}}{\partial(\nabla_X\phi)}\cdot\nabla_X\delta\phi\,\mathrm{d}V \\
&+\int_{\mathcal{B}_0}
\frac{\partial \Psi_{\mathrm{int}}}{\partial\phi}\,\delta\phi\,\mathrm{d}V =0.
\end{aligned}
\end{equation}

Introducing the first Piola--Kirchhoff stress and the conjugate quantities for the non-local field,
\begin{equation}
\mathbf{P}:=\frac{\partial \Psi_{\mathrm{int}}}{\partial \mathbf{F}},
\qquad
\mathbf{Y}:=\frac{\partial \Psi_{\mathrm{int}}}{\partial(\nabla_X\phi)},
\qquad
Y:=-\frac{\partial \Psi_{\mathrm{int}}}{\partial\phi},
\end{equation}
these stationarity conditions can be written in compact form as

\begin{equation}
\begin{aligned}
\delta_{\varphi}\Pi
={}&\int_{\mathcal{B}_0}\mathbf{P}:\nabla_X\delta\varphi\,\mathrm{d}V \\
&-\int_{\mathcal{B}_0}\bar{\mathbf{B}}\cdot\delta\varphi\,\mathrm{d}V
-\int_{\partial\mathcal{B}_0}\bar{\mathbf{T}}\cdot\delta\varphi\,\mathrm{d}A =0,
\end{aligned}
\label{eq:weak_form_u}
\end{equation}

\begin{equation}
\begin{aligned}
\delta_{\phi}\Pi
={}&\int_{\mathcal{B}_0}
\mathbf{Y}\cdot\nabla_X\delta\phi\,\mathrm{d}V
-\int_{\mathcal{B}_0}
Y\,\delta\phi\,\mathrm{d}V =0.
\end{aligned}
\label{eq:weak_form_phi}
\end{equation}

These two weak forms are coupled through the shared free energy density introduced in Eq.~\eqref{eq:internal_energy}, in particular through the local damage variable $\kappa$. Applying integration by parts and localization to the mechanical and non-local statements then yields the coupled strong forms in the reference configuration. For brevity, the derivation is omitted and can be found in \citep{waffenschmidt_gradient-enhanced_2014} and \citep{ostwald_implementation_2019}. The resulting strong forms are given by

\begin{equation}
\begin{aligned}
\nabla_X\cdot\mathbf{P}+\bar{\mathbf{B}}=\mathbf{0}
\quad &\text{in } \mathcal{B}_0,\\
\varphi=\bar{\varphi}
\quad &\text{on } \partial\mathcal{B}_0^{u},\\
\mathbf{P}\mathbf{N}=\bar{\mathbf{T}}
\quad &\text{on } \partial\mathcal{B}_0^{T},
\end{aligned}
\label{eq:strong_form_nloc}
\end{equation}
\begin{equation}
\begin{aligned}
\nabla_X\cdot\mathbf{Y}+Y=0
\quad &\text{in } \mathcal{B}_0,\\
\mathbf{Y}\cdot\mathbf{N}=0
\quad &\text{on } \partial\mathcal{B}_0.
\end{aligned}
\label{eq:strong_form_mech}
\end{equation}

\subsection{Local Damage Evolution}\label{sec:damage_evol_analytic}
Equations \eqref{eq:strong_form_mech} and \eqref{eq:strong_form_nloc} are coupled through the local damage variable $\kappa$, which, locally, scales the strain energy density and thus describes the dissipation of the elastic energy, while also serving as a source term that drives the evolution of the global damage field. To complete the coupled formulation, we need to specify an evolution equation for $\kappa$.

The driving force for the local damage evolution is then defined as the thermodynamic force conjugate to $\kappa$ from the overall free energy by

\begin{equation}
g(\mathbf{F},\phi,\nabla_X\phi,\kappa)
:=-\frac{\partial \Psi_{\mathrm{int}}(\mathbf{F},\phi,\nabla_X\phi,\kappa)}{\partial \kappa}.
\end{equation}
However, it is actually convenient to introduce a specific form of the degradation function
\begin{equation}
f_d(\kappa) = 1-d(\kappa).
\label{eq:damage_function}
\end{equation}
With $d(\kappa)$ a change of variables used in the degradation function because of its physical interpretation. This variable is in the range  $d\in[0,1)$, such that $d=0$ leaves the strain energy unscaled and describes the absence of damage. As damage accumulates, the extreme case of $d\to1$ leads to vanishing strain energy stored regardless of the deformation and constitutes material failure. Thus, $d(\kappa)$ is ideal for the degradation function. However, we still keep $\kappa$ as the primary internal variable because it is unconstrained and it keeps track of the history of the yield surface as will be seen soon. 

The corresponding conjugate variable to $d$ is derived from the free energy 

\begin{equation}
\begin{aligned}
q(\mathbf{F},\phi,\nabla_X\phi,\kappa)
:={}&-\frac{\partial \Psi_{\mathrm{int}}(\mathbf{F},\phi,\nabla_X\phi,\kappa)}{\partial d} \\
={}& -\frac{\partial \psi_{\mathrm{loc}}(\mathbf{F},\kappa)}{\partial d}
   -\frac{\partial \psi_{\mathrm{nloc}}(\mathbf{F},\phi,\nabla_X\phi,\kappa)}{\partial d} \\
={}& q_{\mathrm{loc}}(\mathbf{F},\kappa)
+ q_{\mathrm{nloc}}(\mathbf{F},\phi,\nabla_X\phi,\kappa).
\label{eq:q_q_loc_q_nloc}
\end{aligned}
\end{equation}
Note that, even though we have introduced the auxiliary variable $d$, ultimately we want the explicit dependence on the deformation gradient $\mathbf{F}$, the internal variable $\kappa$, and the global damage field $\phi$. By the choice of $\Psi_{\mathrm{int}}$ and the degradation function in eq. \eqref{eq:damage_function}, the local part of the driving force conjugate to $d$ is nothing more than the strain energy density of the undamaged material.

\begin{equation}
q_{\mathrm{loc}}(\mathbf{F},\kappa) = {\psi_e}(\mathbf{C}) \, .
\end{equation}

For the nonlocal term, we rely on the chain rule of the one-to-one relation $d(\kappa)$ ,
\begin{equation}
\begin{aligned}
q_{\mathrm{nloc}}(\mathbf{F},\phi,\nabla_X\phi,\kappa)
&:=-\frac{\partial \psi_{\mathrm{nloc}}(\mathbf{F},\phi,\nabla_X\phi,\kappa)}{\partial d} \\
&= -\frac{\partial \psi_{\mathrm{nloc}}}{\partial \kappa} \frac{\partial \kappa}{\partial d} \\
&= g(\mathbf{F},\phi,\nabla_X\phi,\kappa)\, \frac{\partial \kappa}{\partial d}.
\end{aligned}
\end{equation}

For the damage evolution, we assume the existence of a dissipation potential $\Phi_d$ defining a yield surface,
\begin{equation}
\Phi_d(\mathbf{F},\phi,\nabla_X\phi,\kappa)
= \mathcal{G}(q(\mathbf{F},\phi,\nabla_X\phi,\kappa))-\kappa \le 0.
\label{eq:potential_function}
\end{equation}
Here it can be seen that $\kappa$ is the history variable that keeps track of the yield surface $ \mathcal{G}(q)$, such that when the yield function $ \mathcal{G}(q)$, evaluated at the conjugate force $q$, is less than $\kappa$, then we are inside of the yield surface, $\Phi_d<0$, and the deformation is elastic, i.e. no damage accumulates. When the driving force $q$ is such that it reaches the yield surface specified by $\kappa$, i.e. $ \mathcal{G}(q)=\kappa$, then $\Phi_d=0$ and damage accumulates according to the associative flow rule
\begin{equation}
\dot{\kappa}=\Delta \lambda\,\frac{\partial \Phi_d}{\partial q}=\Delta \lambda\,\frac{\partial \mathcal{G}}{\partial q},
\label{eq:flow_rule}
\end{equation}
subject to the Karush--Kuhn--Tucker (KKT) conditions
\begin{equation}
\Delta \lambda\ge 0,\qquad \Phi_d\le 0,\qquad \Delta \lambda\,\Phi_d=0.
\label{eq:KKT}
\end{equation}

\section{JAX-FEM Implementation}
The continuum formulation above is implemented in the Python package JAX-FEM \citep{xue2023jax}, a recently developed finite-element package using the JAX library. A major advantage of this framework is the ability for automatic differentiation, which simplifies the implementation of complex constitutive models. For example, automatic differentiation allows for automatic computation of the stress and driving forces given the expression for the free energy. Additionally, another major advantage of this framework is that option to introduce constitutive models which are not closed-form but instead implemented with physics-augmented neural networks using JAX. In this section, we will detail our implementation within the package and the formulation of the coupled damage problem. The central computational idea is that once a scalar free energy density has been defined, automatic differentiation supplies the stresses, non-local fluxes, and local driving forces required by the weak forms. New gradient-enhanced damage models can therefore be introduced by redefining the energy potential rather than by re-deriving the full element residual and constitutive tangent by hand.

\subsection{Residual Formulation}

To implement the coupled damage problem, JAX-FEM requires the specification of a kernel function which specifies how the degrees of freedom are mapped to the residual contributions associated with the weak forms in Eqs.~\eqref{eq:weak_form_u} and \eqref{eq:weak_form_phi}. The finite element discretization, numerical quadrature, and global assembly are provided directly by the package. Here we will specify how to construct the element-wise residual contributions inside the kernel. The quantities provided by JAX-FEM are summarized in Table~\ref{tab:kernel_inputs}. From these we will show how to construct the element-wise residual contributions inside the kernel.

\begin{table}[t]
\centering
\caption{Finite-element quantities passed to the element kernel by JAX-FEM. These are used to construct the element-wise residual contributions inside the kernel.}
\label{tab:kernel_inputs}
\setlength{\tabcolsep}{4pt}
\renewcommand{\arraystretch}{1.1}
\begin{tabularx}{\columnwidth}{@{}>{\raggedright\arraybackslash}p{0.36\columnwidth}X@{}}
\toprule
\textbf{Quantity} & \textbf{Description} \\
\midrule
$\{\mathbf{u}_a\}_{a=1}^{n_{\mathrm{node}}}$ & Displacement nodal values for the current element \\
$\{\phi_a\}_{a=1}^{n_{\mathrm{node}}}$ & Non-local damage nodal values for the current element \\
$\mathbf{X}_q$ & Quadrature-point coordinates \\
$\nabla_X N_a(\mathbf{X}_q)$ & Shape-function gradients \\
$(\nabla_X \delta \mathbf{u},\, \nabla_X \delta \phi)\, J_q w_q$ & Weighted test-function gradients \\
$J_q w_q$ & Quadrature measure \\
$\kappa_q$ & Quadrature history variable \\
\bottomrule
\end{tabularx}
\end{table}

For displacement, the deformation gradient is constructed from the displacement degrees of freedom as

\begin{equation}
\nabla_X \mathbf{u}_q=\sum_{a=1}^{n_{\mathrm{node}}}\mathbf{u}_a\otimes \nabla_X N_a(\mathbf{X}_q),
\qquad
\mathbf{F}_q=\mathbf{I}+\nabla_X \mathbf{u}_q.
\end{equation}
Now with the deformation gradient, the Piola--Kirchhoff stress is constructed as
\begin{equation}
\mathbf{P}_q=\frac{\partial \Psi_{\mathrm{int}}(\mathbf{F}_q,\phi_q,\nabla_X\phi_q,\kappa_q)}{\partial \mathbf{F}_q},
\label{eq:PK1_stress_calc}
\end{equation}
with $\mathbf{P}_q$ being calculated implicitly using automatic differentiation. Note that within the implementation, the damage function is specified as part of the strain-energy density, and is already taken into consideration during the computation of the stress. The residual then can be computed directly as

\begin{equation}
r_{u,a}^{e}=\sum_{q=1}^{n_q}\mathbf{P}_q\,\nabla_X N_a(\mathbf{X}_q)\,J_q w_q,
\end{equation}
so that the displacement element residual vector is
\begin{equation}
\mathbf{R}_{u}^{e}
=
\left[
r_{u,a}^{e}
\right]_{a=1}^{n_{\mathrm{node}}}.
\end{equation}

Similarly for the non-local damage field, the quadrature point values are constructed as

\begin{equation}
\phi_q=\sum_{a=1}^{n_{\mathrm{node}}} N_a(\mathbf{X}_q)\,\phi_a,
\qquad
\nabla_X \phi_q=\sum_{a=1}^{n_{\mathrm{node}}}\phi_a\,\nabla_X N_a(\mathbf{X}_q).
\end{equation}

The quadrature-point internal history variable $\kappa_q$ is introduced here only as an input to the constitutive response. However, the treatement of the history variable in the implementation strategy is not negligible and will be detailed in a later section. 

Using the definitions from Section 2, the non-local conjugate quantities are
\begin{equation}
\mathbf{Y}_q=\frac{\partial \psi_{\mathrm{nloc}}^{\mathrm{grad}}(\nabla_X\phi_q;\mathbf{F}_q)}{\partial (\nabla_X \phi_q)},
\qquad
Y_q=-\frac{\partial \psi_{\mathrm{nloc}}^{\mathrm{plty}}(\phi_q,\kappa_q)}{\partial \phi_q},
\label{eq:nonlocal_conjugate_calc}
\end{equation}

with both terms again being calculated implicitly using automatic differentiation. The gradient contribution to node $a$ is
\begin{equation}
r_{\phi,a}^{e,\nabla}
=
\sum_{q=1}^{n_q}\mathbf{Y}_q\cdot\nabla_X N_a(\mathbf{X}_q)\,J_q w_q,
\end{equation}
while the penalty contribution is
\begin{equation}
r_{\phi,a}^{e,\mathrm{p}}
=
-\sum_{q=1}^{n_q}Y_q\,N_a(\mathbf{X}_q)\,J_q w_q.
\end{equation}
The non-local residual at node $a$ is then
\begin{equation}
R_{\phi,a}^{e}
=
r_{\phi,a}^{e,\nabla}+r_{\phi,a}^{e,\mathrm{p}},
\end{equation}
so that the scalar-field element residual vector is
\begin{equation}
\mathbf{R}_{\phi}^{e}
=
\left[
R_{\phi,a}^{e}
\right]_{a=1}^{n_{\mathrm{node}}}.
\end{equation}
Finally, the element residual returned by the kernel is
\begin{equation}
\mathbf{R}^{e}
=
\begin{bmatrix}
\mathbf{R}_{u}^{e}\\
\mathbf{R}_{\phi}^{e}
\end{bmatrix}.
\end{equation}

\subsection{Local Damage Evolution}\label{sec:damage_evol_analytic_NR}

The computational treatment of the local damage variable follows from assessing the potential function defined in Eq.~\eqref{eq:potential_function}. In the majority of cases, this will be a highly non-linear function and require a Newton-Raphson solver to solve for the update. 
The nonlinear equations that require solution are those coming from the KKT conditions (\ref{eq:KKT}),
\begin{equation}
\begin{aligned}
&{\kappa _{n + 1}} - {\kappa _n} - \Delta \lambda {\left. {\frac{{d\mathcal{G}}}{{dq}}} \right|_{{\kappa _{n + 1}}}} = 0,\\
&{\Phi _d}({\mathbf{F}},\phi ,{\nabla _X}\phi ,{\kappa _{n + 1}}) = 0.
\end{aligned}  
\label{eq:dd_7_no_nn}
\end{equation}
which should be solved for the new value of the history variable $\kappa_{n+1}$ and the Lagrange multiplier $\Delta \lambda$.



As mentioned above, the treatment of the history variable is important for the implementation. Due to the automatic differentiation calculations for~\eqref{eq:PK1_stress_calc} and \eqref{eq:nonlocal_conjugate_calc}, the tangent contributions are also automatically propagated through by JAX-FEM. While this method is extremely useful and reduces the overhead required to implement a damage model, selecting how to update the history variable is important for the accuracy of the solution.

One approach is to use a staggered solution scheme, where the history variable is advanced separately from the global nodal solve. This is done by first solving for the displacement and non-local damage fields, then iterating through quadrature points to perform the update. In the authors experience, this approach resulted in fast convergence with stable global solves and is adequate for problems which are not tightly coupled. However, this approach leads to small errors proportional to the penalty term coupling local, global damage fields and time incrementation. In other words, because the staggered approach introduces a lag between the global and local solves, the non-local driving force in each update is based on the previous step rather than the current step. As a result, for large penalty, there is a risk of drifting away from the true solution, particularly at large time steps.

To avoid this, a monolithic solution scheme can be employed by updating the internal variable for each global Newton iteration. Tangent contributions from $\kappa$ are then added to the global solve, and the update rule given in Eq.~\eqref{eq:dd_7_no_nn}. As will be shown in the results, this approach is more accurate and is preferred but comes with some additional considerations. First, because of the automatic differentiation treatment of the potential function, convergence for the local Newton-Raphson scheme may require a large number of iterations to converge. Secondly, this can create an instability in the global solve if the given $u$ and $\phi$ fields are a poor guess, as at high damage states gradients may become unstable and produce NaNs. 

For the numerical examples shown, we solved directly via PETSc's LU preconditioner in pre-only mode. Each solve was initialized with the last converged solution as the initial guess, with discrete quasi-static loading applied in incremental steps. The local damage evolution utilized an arc-length approach to remain stable. Non-trivial meshes were generated using ABAQUS to produce an input file for JAX-FEM.

\section{Constitutive Models}
The preceding formulation is independent of a particular constitutive law for both local and non-local terms of the strain-energy density. For comparison, we show both closed-form and neural network models to highlight the advantage of the JAX-FEM damage framework. We follow the closed form model provided by \citet{ostwald_implementation_2019}, and additionally propose a data-driven model based on physics-augmented neural networks which is in line with the work of \citet{amiri-hezaveh_physics-augmented_2025}. This section first introduces the closed-form model used in the present benchmark calculations and then outlines how the same framework extends to data-driven strain-energy densities and dissipation potentials.

\subsection{Closed-form Damage Model}
For the closed-form calculations, the undamaged hyperelastic response is described by a compressible Neo-Hookean strain-energy density,
\begin{equation}
\begin{aligned}
\hat{\psi}_{\mathrm{NH}}(\mathbf{C})
&=\frac{\mu_e}{2}(I_1-3)-\mu_e\ln J \\
&\quad+\frac{\lambda_e}{2}(\ln J)^2, \\
I_1&=\mathrm{tr}(\mathbf{C}),\, J=\sqrt{\det \mathbf{C}},
\end{aligned}
\end{equation}
where $\mu_e$ and $\lambda_e$ denote the shear and bulk moduli of the undamaged material. The local elastic contribution then takes the form

\begin{equation}
\psi_{\mathrm{loc}}(\mathbf{F},\kappa)
=f_d(\kappa)\,\hat{\psi}_{\mathrm{NH}}(\mathbf{C}),
\end{equation}
with the degradation function chosen as
\begin{equation}
f_d(\kappa)=\exp\!\left(-\eta_d\langle \kappa-\kappa_d\rangle_+\right)=1-d.
\label{eq:degradation_function}
\end{equation}
where $\eta_d$ and $\kappa_d$ are the damage saturation and threshold parameters, $\langle \cdot \rangle_+$ denotes the Macaulay brackets. The non-local terms are specified then as
\begin{equation}
\psi_{\mathrm{nloc}}^{\mathrm{grad}}(\nabla_X\phi;\mathbf{F})
=\frac{c_d}{2}\,\nabla_X\phi\cdot\mathbf{C}^{-1}\cdot\nabla_X\phi,
\end{equation}
\begin{equation}
\psi_{\mathrm{nloc}}^{\mathrm{plty}}(\phi,\kappa)
=\frac{\beta_d}{2}(\phi-\kappa)^2
\end{equation}
where $c_d$ and $\beta_d$ are the gradient and penalty parameters describing the regularization of the two fields. 

As for the local damage evolution, following the derivation for the potential function from \citet{ostwald_implementation_2019}, the loading function is given by
\begin{equation}
\Phi_d(\mathbf{F},\phi,\nabla_X\phi,\kappa)
=\hat{\psi}_{\mathrm{NH}}(\mathbf{C},J)
+\gamma_d\,\frac{\beta_d(\phi-\kappa)}{\eta_d\,f_d(\kappa)}
-\kappa \le 0.
\label{eq:ostwald_potential}
\end{equation}

\subsection{Physics-Augmented Neural Network Model}
In this section, we generalize the framework for the case of data-driven constitutive equations. First, we briefly discuss the local damage model and subsequently extend the aforementioned analytical nonlocal formulation to the data-driven one.  
\subsubsection{Local Model}
 To define a valid data-driven constitutive equation, there are several mathematical and physical requirements, including objectivity, normality, polyconvexity, and growth conditions (see \cite{amiri-hezaveh_physics-augmented_2025}). In addition to that, the Clausius-Duhem inequality (C-D condition)  ensures non-negativity of the net production of entropy (see \cite{amiri-hezaveh_physics-augmented_2025}). To impose these conditions, firstly, we express the Helmholtz local free energy as follows:
\begin{equation}
\begin{aligned}
&\psi(\mathbf{F},\kappa)=f_d(\kappa)\left(
\mu_e\,\tilde\psi_{iso}(I_{1G}, I_{2G})+\lambda_e\,\psi_{vol}(J)\right), \\
&\tilde\psi_{iso}(I_{1G}, I_{2G})=\psi_{iso}(I_{1G}, I_{2G})-\psi_{iso}(3.0, 3.0),\\
&\psi_{vol}(J)=\left(J + J^{-1} - 2\right)^2,\\
&{I_{1G}} = {J^{ - 2/3}}{I_1},{I_{2G}} = \frac{{{I_2}^3}}{{9{J^4}}},\\
&{I_1} = \operatorname{tr} ({\mathbf{C}}),\,\,{I_2} = \operatorname{tr} (\operatorname{cof} {\mathbf{C}}),
\end{aligned}  
\label{eq:dd_0}
\end{equation}
where the Helmholtz energy is split into an isochoric and a volumetric part. The parameter $\lambda_e$ directly represents a bulk modulus response, just as in the neo-Hookean case, whereas the parameter $\mu_e$ is used here to keep this model as a generalization of the neo-Hookean energy. However, this parameter should only be interpreted as a shear modulus under a specific choice of $\psi_{iso}$; instead, it should just be seen as a factor with units of stress that scales a non-dimensional energy function described by a neural network. As discussed in \cite{linden2023neural}, the volumetric term satisfies the growth condition. Also, $I_{1G}$ and $I_{2G}$ are polyconvex invariants \cite{schroder2003invariant}, and thus, a sufficient condition to fulfill the polyconvexity is to consider these invariants as the inputs of the input convex neural network (ICNN) proposed by \cite{amos2017input}. 
Also, it is straightforward to verify that the energy normality condition is satisfied identically for any ICNN used as $\psi_{iso}$. For the stress normality condition, we note that:
\begin{equation}
\begin{aligned}
&\frac{\partial \psi}{\partial \mathbf{C}}=f_d(\kappa)\left(\mu_e
\frac{\partial \psi_{iso}}{\partial \mathbf{C}}+\lambda_e\frac{\partial \psi_{vol}}{\partial \mathbf{C}}\right), \\
&\frac{\partial \psi_{iso}}{\partial \mathbf{C}}=\frac{\partial \psi_{iso}}{\partial I_{1G}}\frac{\partial I_{1G}}{\partial \mathbf{C}}+\frac{\partial \psi_{iso}}{\partial I_{2G}}\frac{\partial I_{2G}}{\partial \mathbf{C}},\\
& \frac{\partial \psi_{vol}}{\partial \mathbf{C}}=\lambda_e\left(J +J^{-1} - 2\right)\left(1 - J^{-2}\right)J \mathbf{C}^{-1},\\
&\frac{\partial I_{1G}}{\partial \mathbf{C}}=I_3^{-1/3}\left(\mathbf{I}- \frac{1}{3} I_1\mathbf{C}^{-1}\right),\\
&\frac{\partial I_{2G}}{\partial \mathbf{C}}=\frac{I_2^2}{9 I_3^2}\left(3 I_1 \mathbf{I}- 3 \mathbf{C}- 2 I_2 \mathbf{C}^{-1}\right).
\end{aligned}  
\label{eq:dd_1}
\end{equation}
According to $\eqref{eq:dd_1}_{3-5}$, it can be observed that the present formulation identically fulfills the stress normality condition ${\left. {{\mathbf{S}} = 2\frac{{\partial \psi }}{{\partial {\mathbf{C}}}}} \right|_{{\mathbf{C}} = {\mathbf{I}}}} = {\mathbf{0}}$. 

The damage evolution governed exactly as above, in Eqs. (\ref{eq:potential_function})-(\ref{eq:KKT}). The only condition in the model to satisfy the C-D condition is that $\mathcal{G}(q)$ should be an increasing function. Thus, the function $\mathcal{G}(q)$ is described by

\begin{equation}
    \mathcal{G}(q) = \mathcal{N}_\kappa()
    \label{eq:Gq_NN}
\end{equation}

where \textcolor{blue}{$\mathcal{N}_{\kappa}$} is an increasing function with respect to the input argument 
In the local model, $q=q_{loc}$ and there is no contribution from the global damage field $\phi$. Other than that, as just stated, the local damage evolution follows Eqs. (\ref{eq:potential_function})-(\ref{eq:KKT}). The rationale for introducing first the local version of the model, ignoring $\phi$, is that one typically has access to stress-strain curves exhibiting damage under homogeneous conditions. These data can be used to train the neural networks $\psi_{iso}$ and $\mathcal{N}_\kappa$, without any consideration of $\phi$. The learned model can then be incorporated into the finite element implementation for the prediction of non-uniform stress and damage fields.


\subsubsection{Nonlocal Model}
Having defined the local data-driven model, which uses an ICNN for $\psi_{iso}$ and an increasing function for $\mathcal{N}_\kappa$, we state the data-driven nonlocal damage based on the equations in the sections \ref{sec:damage_anlytic}, \ref{sec:damage_evol_analytic}, and \ref{sec:damage_evol_analytic_NR}. The setup is exactly as already described. Nonetheless, to point to specific considerations in the implementation, we note that the Helmholtz free energy \eqref{eq:internal_energy} with the explicit mention if the ICNN strain energy becomes
\begin{equation}
\begin{aligned}
\Psi_{\mathrm{int}}(\mathbf{F},\phi,\nabla_X \phi,\kappa) 
  &= f_d(\kappa)\left(
\mu_e\,\tilde\psi_{iso}(I_{1G}, I_{2G})+\lambda_e\,\psi_{vol}(J)\right) \\
  &\quad + \psi_{\mathrm{nloc}}^{\mathrm{grad}}(\nabla_X\phi;\mathbf{F}) \\
  &\quad + \psi_{\mathrm{nloc}}^{\mathrm{plty}}(\phi,\kappa). 
\end{aligned}  
\label{eq:dd_3}
\end{equation}

The non-local contributions lead to the local damage driving force having two terms, $q=q_{loc}+q_{nloc}$ as stated in Eq. (\ref{eq:q_q_loc_q_nloc}). This driving force, with its local and non-local contributions, then enters the data-driven dissipation potential and flow rule 

\begin{equation}
\begin{aligned}
&{\kappa _{n + 1}} - {\kappa _n} - \Delta \lambda {\left. {\frac{{d\mathcal{N}_\kappa}}{{dq}}} \right|_{{\kappa _{n + 1}}}} = 0,\\
&{\Phi _d}({\mathbf{F}},\phi ,{\nabla _X}\phi ,{\kappa _{n + 1}}) = \mathcal{N}_\kappa (q)-\kappa_{n+1}=0,
\end{aligned}  
\label{eq:dd_7}
\end{equation}

where we have now made the explicit mention of the physics augmented neural network $\mathcal{N}_\kappa$. 


\section{Numerical Examples}
To highlight the advantages of the proposed framework, we present several numerical examples. The first example focuses on the validation of a local damage evolution with the closed-form model with a single element under uniaxial tension, showing that the local damage evolution achieves the expected behavior. Next, a data-driven model is presented with the same framework, showing the extension to any constitutive model defined using a JAX based framework. Single element checks are also performed on the physics-augmented neural network model. 

The subsequent example shows the mesh independence of the implementation and how a purely local damage evolution can lead to mesh dependent results. The same mesh is also used to illustrate the possible divergence of the staggered and monolithic solution schemes. 

The third subsection shows the non-uniform deformation of a notched plate with a data-driven damage evolution, showing a practical application of the data-driven gradient-enhanced formulation. In particular, we examine the convergence of the approach by applying a sensitivity analysis in terms of the mesh and time step refinements.

\subsection{Local Damage Evolution Validation}
\begin{figure*}
    \centering
    \includegraphics[width=0.8\textwidth]{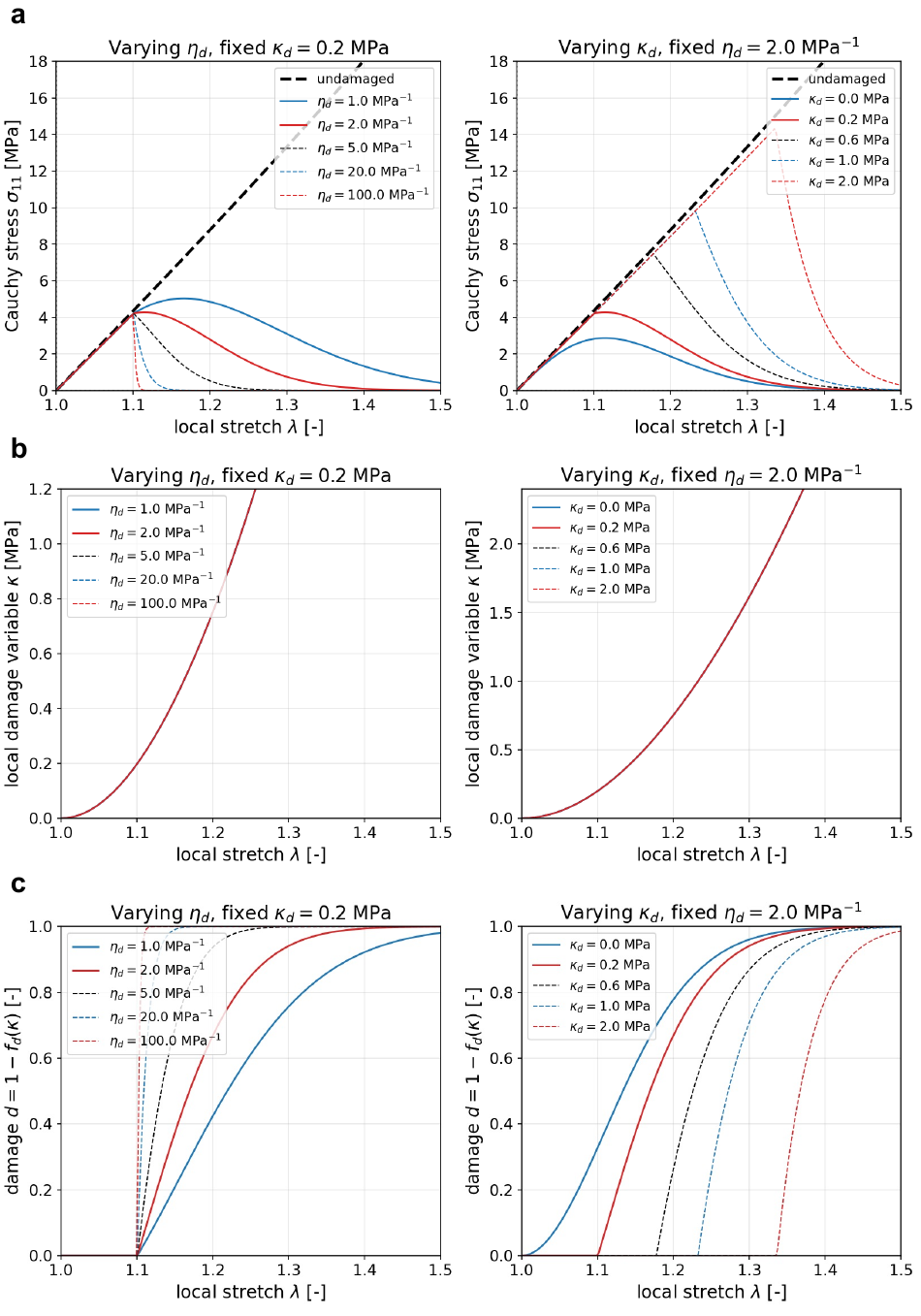}
    \caption{Single-element response under uniaxial loading. a) Axial stress $\sigma_{11}$ vs.\ stretch. b) Local damage variable $\kappa$ vs.\ stretch, c) Damage state $d=1-f_d(\kappa)$ vs.\ stretch. }
    \label{fig:single_element_check}
\end{figure*}

Before analyzing gradient regularization, we first present a validation case for the local evolution in the context of closed-form and data-driven methods. For this analysis, a single $1 \times 1 \times 1\,\mathrm{mm}^3$ HEX8 element is subjected to uniaxial tension. We start with the neo-Hookean material, with elastic parameters chosen as $E=42~\mathrm{MPa}$ and $\nu=0.45$, while for the damage model we consider $\eta_d$ ranging from 1 to 100~$\mathrm{MPa}^{-1}$, $\kappa_d$ ranging from 0 to 2~$\mathrm{MPa}$. To show the impact of the local damage parameters, $\kappa_d$ and $\eta_d$ are varied while holding the other parameter fixed. The element is subjected to a displacement of 0.5 mm, and the response is tracked in terms of the axial component of the stress, the local damage variable $\kappa$, and the damage degradation $d=1-f_d(\kappa)$. Due to the strong softening encountered in this case, the solution is solved using an arc-length continuation solver to avoid numerical instability. Figure~\ref{fig:single_element_check} summarizes the single-element response under uniaxial loading using the three quantities tracked in the benchmark calculations.




For a fixed $\kappa_d$, the axial stress deviates from the virgin neo-Hookean response from the same stretch, with the softening response being a function of $\eta_d$ as expected. Likewise, for a fixed $\eta_d$, varying $\kappa_d$ shows a delayed damage initiation with increasing $\kappa_d$. The evolution of $\kappa$ itself does not change with respect to these parameters. This is because when the non-local regularization is disabled, the local damage variable $\kappa$ is equal to the non-local damage variable $\phi$ causing Eq.~\eqref{eq:ostwald_potential} to reduce exclusively to the neo-Hookean strain-energy density. The damage variable $d$ itself is also a function of $\kappa_d$ and $\eta_d$ as expected, with the damage state being a monotonically decreasing function of $\kappa_d$ and a monotonically increasing function of $\eta_d$ with damage onset occurring for the same $\kappa_d$ and damage evolution showing similar trends for fixed $\eta_d$.

\begin{figure*}
    \centering
    \includegraphics[width=0.8\textwidth]{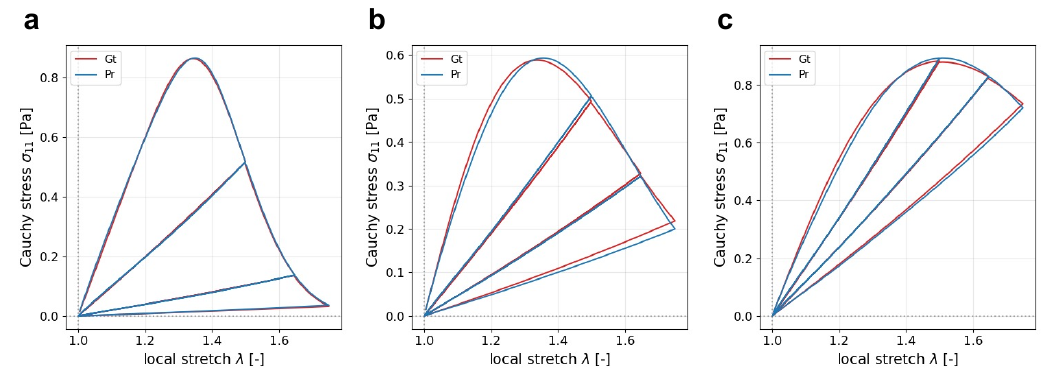}
    \caption{Data-driven model of continuum damage. The same architecture was fitted to three sets of data (a-c), resulting in a  pair of physics-augmented neural networks for each material. }
    \label{fig:data-driven_model_fit}
\end{figure*}

We next fit stress-stretch data showing damage degradation under progressive maximum loading. These data were gathered from published damage models in the literature~\citep{amiri-hezaveh_physics-augmented_2025}. As described in the Methods, the data-driven constitutive model is specified by two ICNNs, one for the isochoric strain energy $\psi_{iso}$ and one for the damage yield function $\mathcal{N}_\kappa$. A pair of networks was trained independently for each of the three materials shown in Fig.~\ref{fig:data-driven_model_fit}. The architecture has sufficient flexibility to capture these three markedly different responses. The material in Fig.~\ref{fig:data-driven_model_fit}a undergoes near-complete failure by $\lambda \approx 1.6$, with the stress dropping to nearly zero by this deformation. The material in Fig.~\ref{fig:data-driven_model_fit}b also shows progressive softening across four loading cycles of increasing amplitude. The material in Fig.~\ref{fig:data-driven_model_fit}c exhibits the least degradation of the three, retaining substantial stiffness even at the largest applied stretch. In all cases, the predicted response (Pr) closely follows the ground truth (Gt), demonstrating that the ICNN architectures for $\psi_{iso}$ and $\mathcal{N}_\kappa$ accurately reproduce the target stress-stretch behavior. 

\subsection{Uniform Deformation and Mesh Independence}
\label{sec:uniform_deformation_and_mesh_independence}

With the established local evolution for the closed-form and data-driven models, we now present the mesh independence of the gradient enhanced formulation. Here, a plate in uniaxial tension is considered, with the plate dimensions of $10\times 10\times 1~\mathrm{mm}^3$. The material parameters are chosen as $E=210~\mathrm{MPa}$ and $\nu=0.3$ for the neo-Hookean material, the local damage parameters chosen as $\eta_d=0.002~\mathrm{MPa}^{-1}$ and $\kappa_d=0.1~\mathrm{MPa}$, and the gradient damage parameters chosen as $c_d=1~\mathrm{MPa}^{-1}~\mathrm{mm}^2$ and $\beta_d=1000~\mathrm{MPa}^{-1}$. The loading is applied by prescribing a horizontal displacement in 1,000 increments on the right boundary while the left side and selected corner points suppress rigid-body motion. 

\begin{figure*}
\centering
\includegraphics[width=0.8\textwidth]{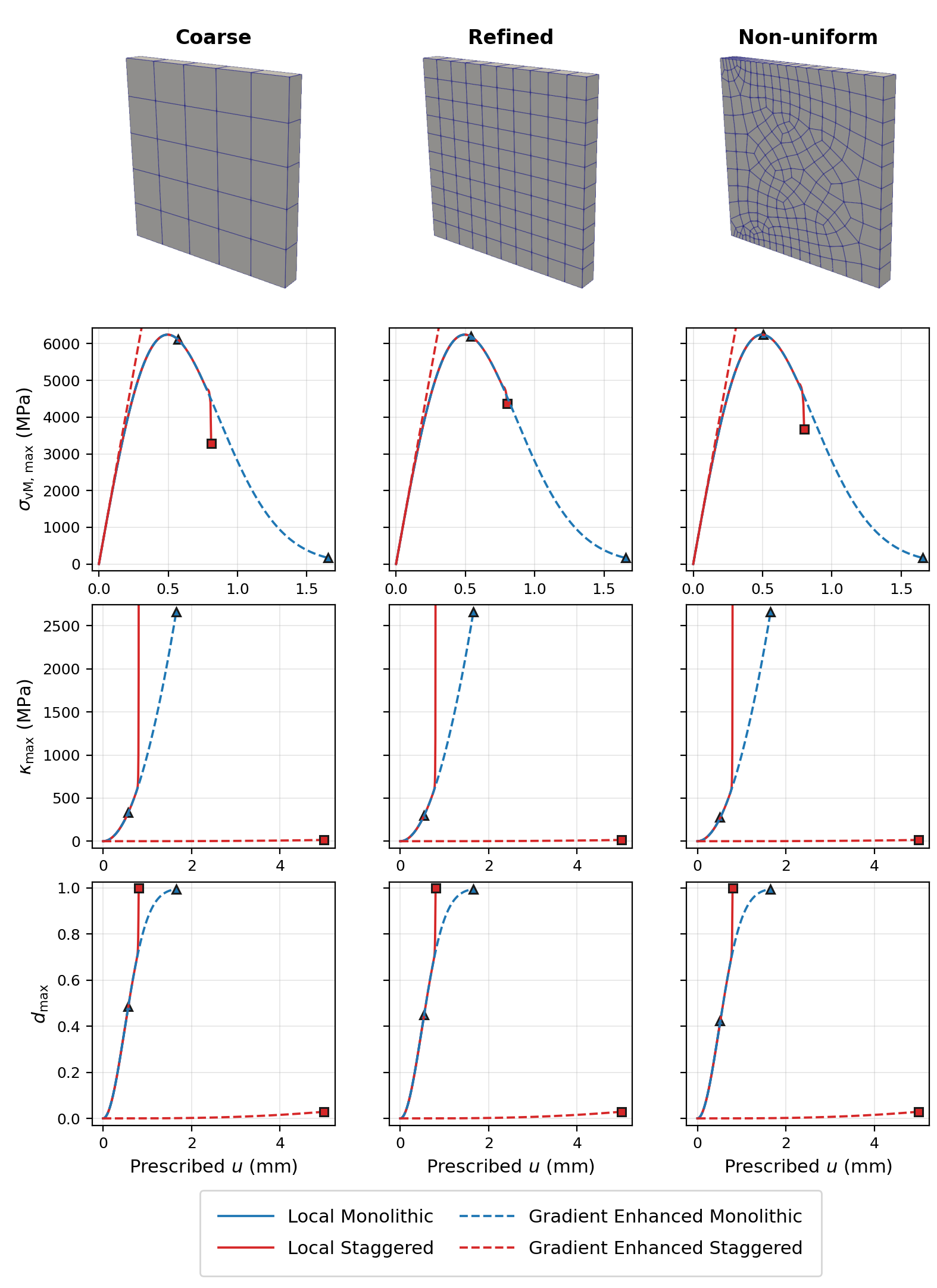}
\caption{Comparison of the staggered and monolithic solution schemes for the uniform deformation and mesh independence study. Simulations terminated when damage variable reached $d=0.995$ or the solver failed to converge.}
\label{fig:uniform_deformation_solstrat}
\end{figure*}

\begin{figure*}
\centering
\includegraphics[width=0.8\textwidth]{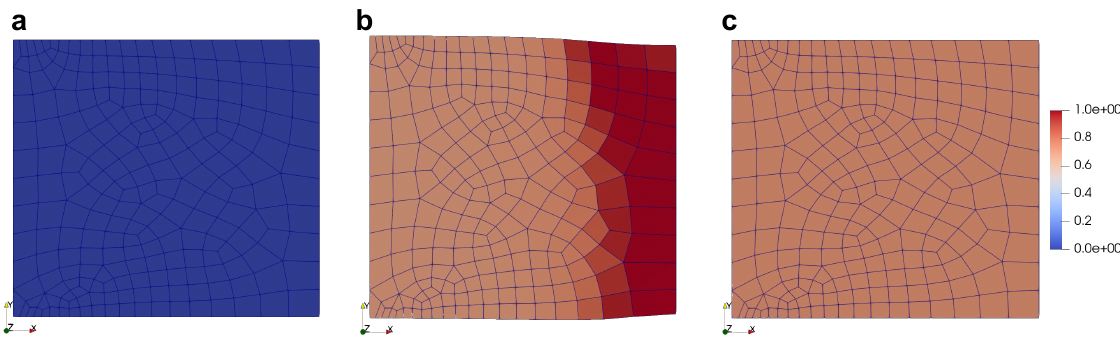}
\caption{Damage field $d$ at a prescribed displacement of $u_x = 0.75 \text{mm}$. Simulations terminated at $d=0.995$. (a) fails to capture damage evolution effectively, (b) shows expected damage but localizes due to mesh selection, (c) avoids localization while capturing expected damage evolution.}
\label{fig:uniform_deformation_damage_fields}
\end{figure*}

Three meshes are compared: a coarse, refined, and a non-uniform mesh. Stress, local damage variable, and damage state are tracked as a maximum value over the element. For each mesh, a pairing of either a local-only or non-local approach with either a staggered or monolithic computational strategy are then compared. The results are summarized in Figure~\ref{fig:uniform_deformation_solstrat}. The comparison of the solver strategy is shown by visualizing the damage variable $d$ for the local staggered (left), gradient-enhanced monolithic (center), and gradient-enhanced staggered (right) formulations at comparable loading in Figure~\ref{fig:uniform_deformation_damage_fields}.

The gradient enhancement problem and the choice of implementing the local damage update plays a significant role in how solutions change, even under a simple example as uniaxial tension. Due to the stiff coupling of the problem, the gradient-enhanced staggered solution quickly diverges from the other strategies and drastically underestimates the damage state. This happens because of the potential function in Eq.~\eqref{eq:ostwald_potential}, which has an overarching contribution from the term containing $\beta_d$, i.e. the penalty term coupling local and global damage variables. As a result, the strain energy function, which should be the main driver for damage accumulation, is unable to induce changes in $\kappa$. In other words, the error in the staggered can decrease with smaller $\beta_d$, as this is the parameter controlling the penalty term in the non-local formulation, but it would induce a significant lag between the local and global damage fields and thus negate the advatange of the gradient-enhanced damage model.

The two remaining approaches show identical damage evolution initially but the local monolithic strategy fails to converge. Softening of the material  results in large Newton-Raphson steps, and with the coupling to the local damage update this causes unrealistic trial states. The purely local staggered approach does perform better in this scenario, but comes with the risk of localization of the damage evolution. This is visualized in Figure~\ref{fig:uniform_deformation_damage_fields} by the presence of a damage band at the right edge showing mesh dependency. By coupling the gradient problem to the monolithic scheme, the simulation pushes past both of these issues and terminates at the prescribed damage threshold. Additionally, the solution does not show any mesh depencency in the damage state, showing the robustness of the gradient-enhanced formulation solved with the monolithic scheme.

\subsection{Notched Plate}
We finish the numerical examples with a benchmark case for a non-trivial geometry. The notched plate example from \citet{ostwald_implementation_2019} is considered, with the plate dimensions of $100 \times 100 \times 1~\mathrm{mm}^3$. Material and damage parameters are those from the data-driven model in Fig. \ref{fig:data-driven_model_fit}b. The constants employed for this example are the same as those used in the previous section, except that $\eta_d=0.001~\mathrm{MPa}^{-1}$ is selected. For the Newton–Raphson procedure used to solve the local nonlinear damage equations, the maximum number of iterations is capped at $100$. 
The loading in this case comes from prescribing a horizontal displacement of $25~\mathrm{mm}$ on the right boundary while the left side and selected corner points suppress rigid-body motion. To show the convergence of the solution, we consider mesh discretization and loading step refinement. Three mesh resolutions are selected: 1) coarse: $241$ elements, 2) medium:$625$ elements, and 3) fine: $2350$ elements. In all cases, the loading step is set to $25$. For the incrementation step sensitivity analysis, the three cases are investigated: 1) $25$ steps, 2) $50$ steps, and 3)  $100$ steps. 

\begin{figure*}
\centering
\includegraphics[width=0.8\textwidth]{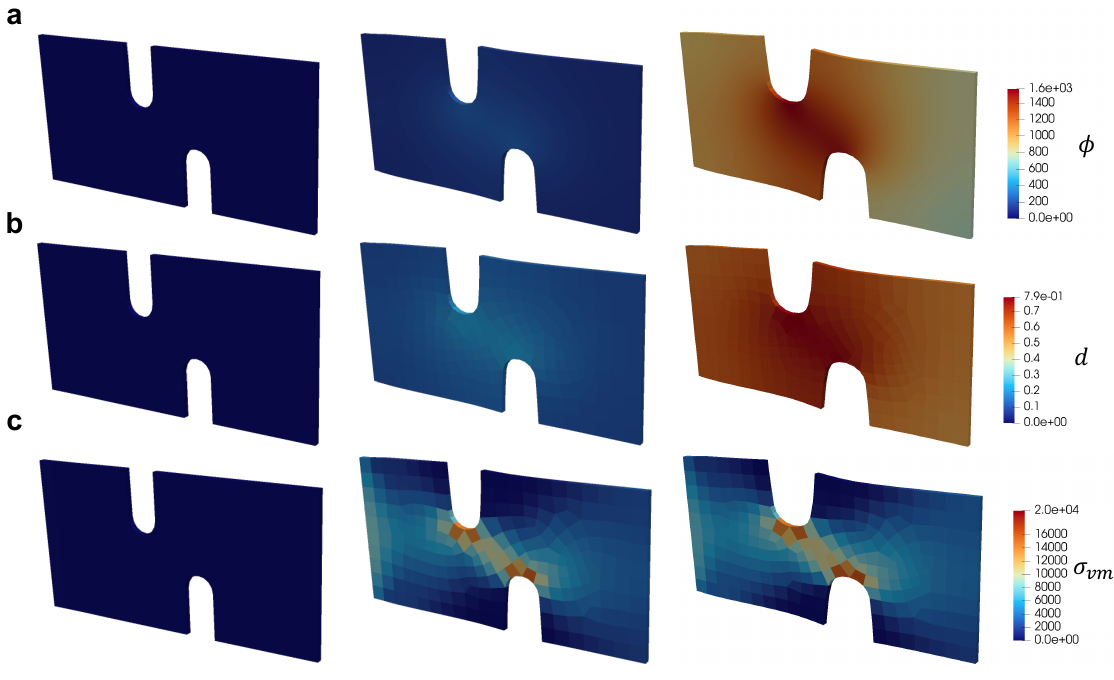}
\caption{Notched plate simulations at three increment frames of the simulation with a final $25~\mathrm{mm}$ displacement on the right boundary. a) Global damage field $\phi$, b) local damage degradation $d$, c) Von Mises stress.}
\label{fig:notched_plate}
\end{figure*}

Figure \ref{fig:notched_plate} shows that the global field $\phi$ evolves smoothly over the domain and this is captured by the finite element interpolation since the field $\phi$ is continuously interpolated based on the nodal values. In contrast, the local damage degradation and stresses are plotted per element in Figure \ref{fig:notched_plate}. Even though the stress shows concentration at the notches, the damage field is diffused due to the PDE governing the field $\phi$, which is coupled through the driving force to the history variable $\kappa$ and, through the degradation function, to the actual damage degradation $d(\kappa)$ depicted in Figure \ref{fig:notched_plate}. This example shows the finite element implementation of the data-driven model based on physics augmented neural networks in a non-trivial three-dimensional simulation. 


\begin{figure*}
\centering
\includegraphics[width=0.8\textwidth]{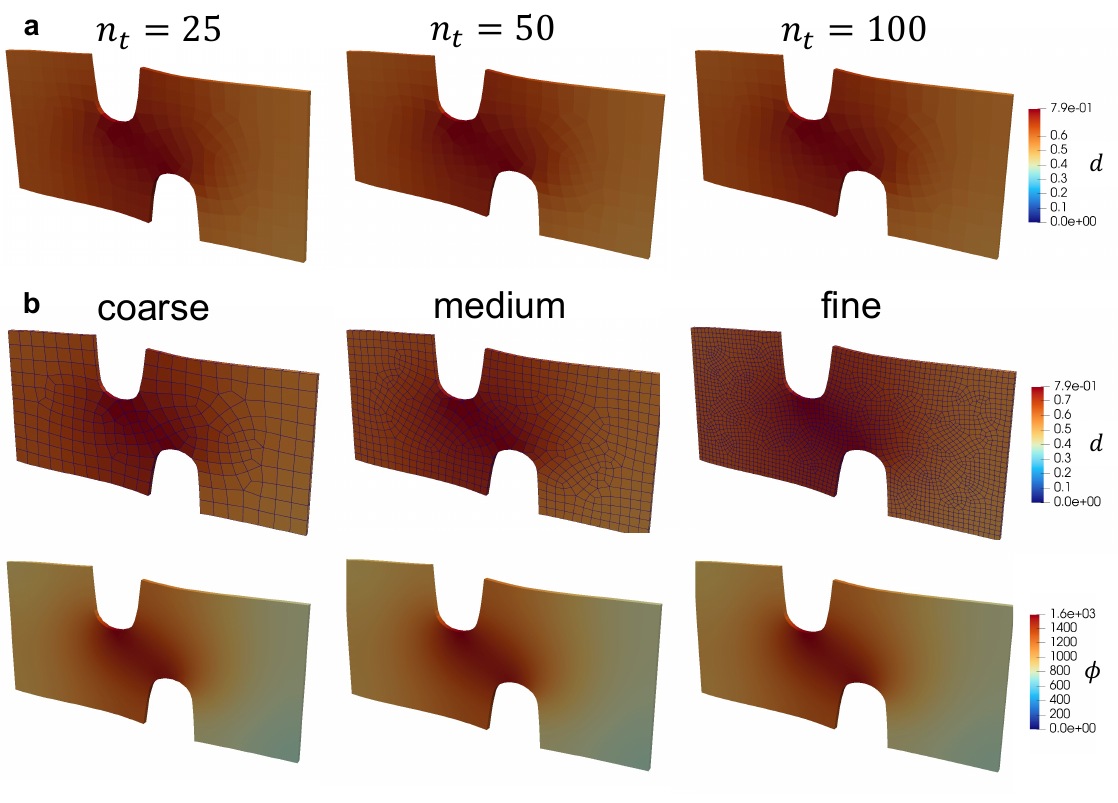}
\caption{Notched plate simulations. a) Convergence with respect to the number of increment steps. b) Convergence with respect to mesh refinement.}
\label{fig:notched_plate_convergence}
\end{figure*}

Figure \ref{fig:notched_plate_convergence} shows the mesh-independence of the solution with respect to both number of solution steps and mesh refinement. In Figure \ref{fig:notched_plate_convergence}a, the notched plate problem is solved with the coarse mesh for either 25, 50, or 100 increments. The damage degradation $d$ is shown, with the same range in the contour plot across all three cases. The fields are identical, showing that the method (monolithic solution of local and global problems) is stable with respect to step size. Figure \ref{fig:notched_plate_convergence}b similarly illustrates the convergence with respect to mesh size. The damage degradation $d$ shows again identical contours across the three meshes. Maximum degradation of $d=0.79$ occurs near the notched region, but the degradation stays diffuse and it does not localize to a single band of element. The reason for the mesh-independent behavior is due to the global field $\phi$, which obeys a diffusion-type PDE. The contour of $\phi$ in Figure \ref{fig:notched_plate_convergence}b show the diffused nature of the regularizing field, which is couple to the yield surface history variable $\kappa$ at the integration-point level as a source term for damage in the driving force $q$.

\section{Discussion and conclusion}

In this work we presented a finite element implementation of a data-driven  gradient-enhanced damage model within the open-source package JAX-FEM. The 
implementation brings together two recent developments in scientific machine learning: physics-augmented neural network constitutive models and differentiable finite element methods. 

Differentiable programming is an increasingly attractive paradigm for scientific  computing because it removes the burden of deriving and implementing analytical derivatives for constitutive model evaluation, tangent construction, and solver sensitivities \citep{sunil2024fe,zhang2022hidenn,vskardova2025finite}. Beyond implementation convenience, the ability to differentiate through the solver itself opens the door to gradient-based design optimization and inverse problems. JAX-FEM, built on the Python library JAX, provides exactly this infrastructure in an open-source setting \citep{xue2023jax}. A particular practical advantage for the present work is the large and growing ecosystem of JAX-based data-driven constitutive modeling frameworks, such as in our previous work \cite{tac2022data,amiri-hezaveh_physics-augmented_2025}.

Data-driven constitutive modeling has advanced rapidly, with neural networks offering unmatched flexibility as universal function approximators \citep{fuhg2025review}. However, unconstrained networks are prone to overfitting and tend to extrapolate poorly outside the training regime. Physics constraints are therefore crucial, and can be imposed either through regularization terms in the loss or, more robustly, directly by architecture design. The latter is usually refer to as physics-augmented neural networks \citep{fuhg2025review}. Building on our previous work~\citep{tac2022data,amiri-hezaveh_physics-augmented_2025}, we represent the elastic energy $\psi_{iso}(I_{1G}, I_{2G})$ and the damage 
yield function $\mathcal{N}_\kappa(q)$ with ICNNs, which enforce polyconvexity and satisfaction of the Clausius-Duhem inequality by construction. These 
networks demonstrate excellent agreement with three datasets drawn from the literature~\citep{amiri-hezaveh_physics-augmented_2025}, and show that a single architecture has sufficient flexibility to accurately describe markedly different softening behaviors while remaining thermodynamically consistent.

While prior work, including our own, has focused primarily on the constitutive modeling framework, numerical implementation is equally important in practice. For damage models in particular, a critical issue is mesh dependence: when only a local damage model is used, strain softening leads to spurious localization that is entirely controlled by the mesh discretization \citep{ostwald_implementation_2019}. Non-local formulations resolve this by coupling the local constitutive response to a spatially regularizing field that introduces a physical length scale. Following closely the gradient-enhanced framework of \cite{ostwald_implementation_2019}, and extending it to nonlinear materials with arbitrary elastic energy and yield functions, we implement a  general gradient-enhanced damage model in JAX-FEM. The implementation is validated through agreement with the local model, a mesh dependence study demonstrating the regularizing effect of the gradient enhancement, and a comparison of staggered versus monolithic solution schemes. A notched plate example demonstrates the  practical utility of the framework in a realistic setting. Together, these contributions provide the community with access to an open-source gradient-enhanced damage with physics-augmented neural network constitutive laws, applicable to a broad class of soft materials.

\section*{Acknowledgments}
The authors gratefully acknowledge support from the U.S. Army Research Office under Award W911NF261A010.

\section*{Data Availability}
\url{https://github.com/mwilkinson1/damage_jaxfem}

\bibliographystyle{cas-model2-names}
\bibliography{refs}

\end{document}